\documentclass[aps,pre,twocolumn,showpacs,floatfix,superscriptaddress]{revtex4-1}
\usepackage{graphicx}
\usepackage{amssymb}
\usepackage{dcolumn}
\usepackage{bm}
\usepackage{dcolumn}
\usepackage{url}

\def\s#1{_{\rm #1} }
\def\sp#1{^{\rm #1} }
\def\Tr#1{{\rm Tr}\left( #1 \right)}
\def\Det#1{{\rm Det}\left( #1 \right)}
\def\Diag#1{{\rm Diag}\left( #1 \right)}
\def\lz{\ell_{\parallel}}
\def\lp{\ell_{\bot}}
\def\tens#1{\underline{\underline{{#1}}}}

\def\e{ {\rm e } }
\def\bea{\begin{eqnarray}}
\def\eea{\end{eqnarray}}
\def \be{\begin{equation}}
\def \ee{\end{equation}}

\begin{document}
\title{Photodynamics of stress in clamped nematic elastomers}
\author{Milo\v{s} Kne\v{z}evi\'{c}}
\email{mk684@cam.ac.uk}
\author{Mark Warner}
%\email{mw141@cam.ac.uk}
\affiliation{Cavendish Laboratory, University of Cambridge, Cambridge CB3 0HE, United Kingdom}
\author{Martin \v{C}opi\v{c}}
\affiliation{Cavendish Laboratory, University of Cambridge, Cambridge CB3 0HE, United Kingdom}
\affiliation{Faculty of Mathematics and Physics, University of Ljubljana, Jadranska 19, SI 1001 Ljubljana, Slovenia}
\author{Antoni S\'{a}nchez-Ferrer}
\affiliation{Institute of Food, Nutrition and Health, ETH Z\"{u}rich, Schmelzbergstrasse 9, 8092, Z\"{u}rich, Switzerland}
\date{\today}

\begin{abstract}
We describe the complex time-dependence of the build up of force exerted by a clamped photo-elastomer under
illumination. Nonlinear (non-Beer) absorption leads to a bleaching wave of a significant \textit{cis} isomer
dye concentration deeply penetrating the solid with a highly characteristic dynamics. We fit our experimental
response at one temperature to obtain material parameters. Force-time data can be matched at all other temperatures
with no fitting required -- our model provides a universal description of this unusual dynamics.
The description is unambiguous since these are clamped systems where gross polymer motion is suppressed as a
possible source of anomalous dynamics. Future experiments are suggested.
\end{abstract}


\maketitle

\section{Introduction}

Unusual properties of liquid crystal elastomers (LCE) arise from a coupling
between the liquid crystalline ordering of mesogenic molecules and the elasticity
of the underlying polymer network. Crosslinked networks of polymer chains of a LCE include
mesogenic units which either belong to the polymer backbone (main-chain LCE) or to
side units pendent to the backbone (side-chain LCE)~\cite{warnerbook:07}.
The shape of a monodomain nematic LCE strongly depends on the temperature dependent
nematic order parameter $Q(T)$. This connection is a consequence of the coupling of $Q(T)$ with
the average polymer chain anisotropy. By increasing the temperature of a LCE, $Q(T)$
decreases which leads to a decrease of the polymer backbone anisotropy. This decrease
of anisotropy manifests itself as an uniaxial contraction of the LCE sample.

Mechanical change can be realized also through a change of nematic order by
other means. As first shown in LCEs by Finkelmann \textit{et al}.~\cite{finkelmann:01}, changes can also be
achieved by introducing photo-isomerizable groups (e.g. azobenzene) into their chemical structure.
These structures will be referred to as nematic photo-elastomers.
By absorbing a photon, azobenzene dye molecules can make transitions, with a quantum efficiency $\eta\s{t}$, from their linear (\textit{trans}) ground state to the excited bent-shaped (\textit{cis}) state.
While the rodlike \textit{trans} molecules contribute to the overall nematic order, the bent
\textit{cis} molecules act as impurities which reduce the nematic order parameter and lower
the nematic-isotropic transition temperature. The illumination of photo-elastomers causes
the reduction of nematic order, which in turn produces an uniaxial contraction of the sample.
On switching off the irradiation, the nematic order parameter recovers its initial dark state
value, which results in a macroscopic expansion of the sample.

The characteristic time of the mechanical response of the sample is in the rather wide range of milliseconds~\cite{white:08,white:10}
to hours~\cite{finkelmann:01,tajbakhsh:01,cviklinski:02,harvey:07,ferrer:11}. To see whether slow on-response in~\cite{finkelmann:01}
was caused by a slow polymer dynamics of the nematic elastomer, or by a slow photo-isomerization kinetics, experiments were
done~\cite{tajbakhsh:01,cviklinski:02,harvey:07} on clamped photo-elastomers.
Contrary to the case of the measurements on a freely suspended sample, the clamped setting
requires no physical movement of chains in the network.
In such settings the sample is fixed at both ends of its longest dimension
(along the nematic director). Then, upon irradiation the sample cannot shrink and a retractive force is exerted
on the clamps -- see Fig.~\ref{fig1} for an illustration. This force is measured as a function of time until the photo-stationary state is reached.
On switching off the irradiation the relaxation of the force begins.
It was first observed by Cviklinski \textit{et al}.~\cite{cviklinski:02} that the dynamics of the nematic order parameter
matches the dynamics of the mechanical response meaning that the rate-limiting process is dominantly the photo-isomerization. They also found that their systems display simple exponential processes for the build up of and decay of force.

Subsequently, it has been shown in other systems~\cite{ferrer:11} that the stress -- temperature experimental data for the on-process
can be approximately fitted to a simple stretched exponential form $\overline{\sigma}\s{n}(t) = 1 - \exp [-(t/\tau\s{on})^{\beta\s{on}}]$, where
$\overline{\sigma}\s{n}(t)$ is the normalized stress exerted on the clamps (stress at time $t$ divided by the photostationary stress),
and $\tau\s{on}$ and $\beta\s{on}<1$ are fit parameters. Similarly, it was found that the normalized stress
in the off-process can be fitted to the nearly exponential law $\overline{\sigma}\s{n}(t) = \exp [-(t/\tau\s{off})^{\beta\s{off}}]$,
where $\beta\s{off} \approx 0.9$. Such complex dynamical response cannot be attributed (as is usual) to polymer dynamics since
we have clamping. In this paper we show that these findings can in fact be successfully described by our simple model
of photodynamics and its conversion into stress in clamped nematic elastomers.

According to the Beer law of light absorption the light propagating in a thick absorbing sample is
attenuated at constant rate. The light intensity at a depth $x$ into the sample is $I(x) = I_0 {\rm e}^{-x / d\s{t}}$,
where $I_0$ is the incident intensity, and $d\s{t}$ is the characteristic penetration depth of a given material due to
absorption by \textit{trans} isomers.
However, it has been shown that the simple Beer law for light attenuation through the sample containing
dye molecules might be inaccurate, due to the so-called photobleaching effect~\cite{statman:03,corbettdyn:03,corbettdyn:02,corbettdyn:01}.
This effect is caused by depletion of \textit{trans} isomers, which allows light to penetrate to greater depths
than those predicted by Beer's law.
If dye molecules that absorbed photons don't return to their \textit{trans} state immediately, the new
photons falling on the sample cannot be absorbed in the initial layers and therefore propagate through
the sample following a nonlinear absorption law.

Using the model described in Sec.~\ref{sect:model}, we calculate the stress exerted on the clamps during and after light irradiation.
Then we relate the stress to the light absorbance $\mathcal{A} = - \ln (I/I_0)$ at the back of the sample. We apply
the full nonlinear absorption model which takes into account the forward \textit{trans} to \textit{cis} and
back \textit{cis} to \textit{trans} photo-isomerization as well as thermal-isomerization of \textit{cis} molecules
back to the \textit{trans} state~\cite{corbettdyn:02}.
The major puzzle in photo-actuation is thereby addressed. Beer penetration depths are typically in the $1-10 ~\mu{\rm m}$
range for normal dye loadings and hence orders of magnitude less than the sample thickness. If a small volume fraction
of solid is photo-contracted, one expects the overall mechanical response to be small in the Beer limit. We show that the
stress is proportional to the \textit{cis} concentration and thus bleaching allows an appreciable sample volume to
contribute to the force as a wave of \textit{trans} to \textit{cis} conversion deeply penetrates.
To test the validity of our model we fit the on-process $\overline{\sigma}\s{n}(t)$ data of Ref.~\cite{ferrer:11}
at one temperature in order to fix the material constants determining the photo-processes. The $\overline{\sigma}\s{n}(t)$ response
at other temperatures then does not need fitting -- the same material constants suffice, after they are
shifted by the separately measured changes in thermal relaxation times with temperature change, to reproduce the
stress response. This remarkable universal agreement between theory and experiment confirms the hypothesis of the
domination of photo-isomerization over polymer dynamics. We hence explain the observed stretched exponential
($\beta\s{on}<1$) processes in terms of nonlinear spatio-temporal photodynamics, rather than the usual (unknown)
polymer relaxation processes normally assumed to be behind such complex dynamics. Our analysis reveals that it is
important to consider back \textit{cis} to \textit{trans} photoconversion.

\section{Model}\label{sect:model}

In this Section we consider a simple model of stress dynamics of clamped nematic photo-elastomers. The
resulting stress is calculated within a nematic rubber model in Section~\ref{subsect:A}, while
the process of nonlinear light absorption is outlined in Section~\ref{subsect:B}.

\subsection{Stress exerted on the clamps}\label{subsect:A}

Long polymer chains have a Gaussian distribution, becoming anisotropic if they contain mesogenic molecules.
The elastic free energy density of a nematic rubber in response to
a deformation gradient tensor $\tens{\Lambda}$ can be expressed as~\cite{warnerbook:07}
\be
F = \frac{1}{2} \mu \Tr {\tens{\ell}\s{0} \cdot \tens{\Lambda}^{\rm T} \cdot \tens{\ell}^{-1} \cdot \tens{\Lambda}}
+ \frac{1}{2} \mu \ln \left ( \frac{\Det{\tens{l}}}{\Det{\tens{l}\s{0}}} \right ),
\label{fenergy}
\ee
where $\mu = n\s{s}k\s{B}T$ is the shear modulus in the isotropic state ($n\s{s}$ is the number of network strands per unit volume).
The tensors $\tens{\ell}$ and $\tens{\ell}\s{0}$ generalize the Flory step length, whence directions parallel and perpendicular
to the nematic director $\underline{n}$ have values $\lz$ and $\lp$. The matrix $\tens{\ell}$ describes the current Gaussian
distribution after deformation $\tens{\Lambda}$, while $\tens{\ell}\s{0}$ gives the initial step lengths. As rubber changes shape
at constant volume $\Det{\tens{\Lambda}} = 1$. Taking $\underline{n}$ along the $z$-axis, $\tens{\ell}$ assumes the diagonal
form $\tens{\ell}= \Diag{\lp,\lp,\lz}$. We adopt a simple freely jointed polymer model, with a step length $a$ in the isotropic state.
The elements of $\tens{\ell}$ are $\lz = a(1+2Q) $ and $\lp = a(1-Q)$, with $Q$ being the nematic order parameter.
Although crude, this model quite accurately describes a wide range of main-chain and side-chain LCEs~\cite{warnerbook:07,finkelmann:01t}.
We are only concerned with derivatives of $F$ with respect to $\Lambda$ and we shall suppress $\Lambda$-independent terms in
Eq.~(\ref{fenergy}).

Consider an elastomer in an initial nematic state at temperature $T$, with $\tens{\ell}\s{0} = \Diag{\lp^0,\lp^0,\lz^0}$.
Illumination changes $\tens{\ell}\s{0}$ to $\tens{\ell}\s{p} = \Diag{\lp\sp{p},\lp\sp{p},\lz\sp{p}}$. A free sample suffers
uniaxial spontaneous photodeformation $\tens{\lambda}\s{p} = \Diag{1/\sqrt{\lambda\s{p}},1/\sqrt{\lambda\s{p}},\lambda\s{p}}$ directed along $\underline{n}=\underline{z}$, with its principal contraction $\lambda\s{p}<1$, and perpendicular elongation $1/\sqrt{\lambda\s{p}}$
due to incompressibility $\Det{\tens{\lambda}\s{p}} = 1$.
The free energy density~(\ref{fenergy}) corresponding to $\tens{\Lambda} = \tens{\lambda}\s{p}$ is
\be
F = \frac{1}{2} \mu \left ( \frac{2}{\lambda\s{p}}\frac{\lp\sp{0}}{\lp\sp{p}} + \lambda\s{p}^2 \frac{\lz\sp{0}}{\lz\sp{p}} \right ).
\label{fenergyphoto}
\ee
Minimisation over $\lambda\s{p}$ gives the spontaneous photocontraction $\lambda\s{ps} = (\lz\sp{p} \lp\sp{0}/\lp\sp{p} \lz\sp{0})^{1/3}$.
Clearly, $\lambda\s{ps}<1$ since $\lz\sp{p}/\lp\sp{p} < \lz\sp{0}/\lp\sp{0}$, that is the sample becomes less anisotropic on illumination.

\begin{figure}
  \includegraphics[width=8.7cm]{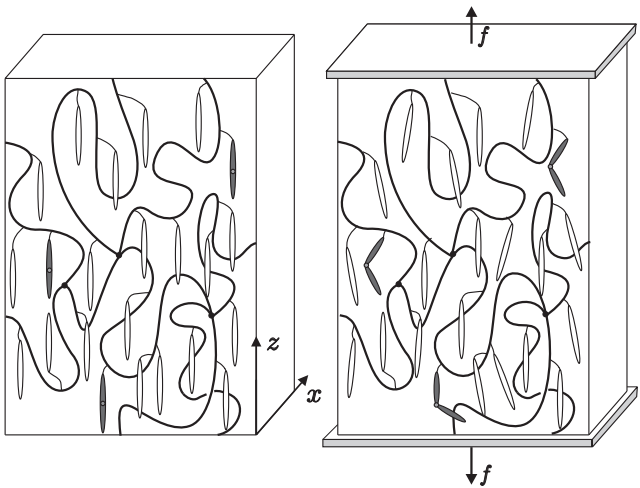}
  \caption{A schematic of a nematic elastomer with regular mesogenic molecules (white rods) and
  photo-actives (gray). In the dark all photo-active molecules are found in their linear \textit{trans} state
  (left panel). Upon illumination some photo-active molecules undergo a transition to the bent \textit{cis} state
  (right panel), which would lead to contraction if the elastomer were freely suspended.
  Such contraction is prevented by the clamps, resulting in a force $f$ on the clamps.
  }
\label{fig1}
\end{figure}

Clamping (see Fig.~\ref{fig1}), in effect stretches the sample by $\lambda = 1/\lambda\s{ps}$ along $\underline{n}$ to restore the sample
to the length before illumination. It is known~\cite{finkelmann:01t} that strain little perturbs the underlying nematic order.
We assume that $\tens{\ell}\s{p}$ is unchanged after this stretching. Taking $\tens{\Lambda} = \tens{\lambda} \cdot \tens{\lambda}\s{ps} = \Diag{1/\sqrt{\lambda \lambda\s{ps}},1/\sqrt{\lambda \lambda\s{ps}}, \lambda \lambda\s{ps}}$ in (\ref{fenergy}) and
the above value for $\lambda\s{ps}$, one gets
\be
F = \frac{1}{2} \mu \left ( \frac{\Det{\tens{\ell}\s{0}}}{\Det{\tens{\ell}\s{p}}} \right )^{1/3} \left ( \lambda^2 + \frac{2}{\lambda} \right ).
\label{fenergylambda}
\ee
The result is the same as that of a classical elastomer with a renormalized shear modulus $\mu$.
If the formation state were isotropic followed by cooling to a nematic before illumination, $F$ is as in
(\ref{fenergylambda}) with a different non-light dependent prefactor~\cite{warnerbook:07}.

We denote the area of the sample in the nematic state before illumination by $A_0$ and its length along the director by $L_0$.
Incompressibility gives $A_0 L_0 = A\s{ps} L\s{ps}$, where $A\s{ps}$ and $L\s{ps}$ are the area and the length of a free sample after illumination.
The force exerted by a photo-elastomer due to the stretching $\lambda$ is
\be
f = A_0 L_0 \frac{\partial F}{\partial (\lambda L\s{ps})} = A\s{ps} \frac{\partial F}{\partial \lambda},
\label{force}
\ee
where $A\s{ps} = A_0 / \lambda\s{ps}$. After differentiation, we take $\lambda= 1/\lambda\s{ps}$ since there is clamping.
The stress $\sigma = f/A_0$ exerted on the clamps takes the form
\be
\sigma = \mu \left ( \frac{\Det{\tens{\ell}\s{0}}}{\Det{\tens{\ell}\s{p}}} \right )^{1/3}
\left ( \frac{1}{\lambda\s{ps}^2} - \lambda\s{ps} \right).
\label{stress}
\ee
As the shape of the sample is not changing upon the illumination, true and engineering stresses are the same.

The step lengths before illumination are $\lz\sp{0}=a(1+2Q(T))$ and $\lp\sp{0}=a(1-Q(T))$. Upon illumination,
the \textit{cis} isomers act as impurities which lower the nematic -- isotropic transition temperature, which can also be seen as a light-dependent increase of the temperature T~\cite{finkelmann:01}. The elements of $\tens{\ell}\s{p}$ are
$\lz\sp{p}=a(1+2Q(T\s{eff}))$ and $\lp\sp{p}=a(1-Q(T\s{eff}))$, where $T\s{eff}$ is a fictitious effective temperature
which is the actual temperature $T$ increased by a light-dependent term $\Delta T$, whence $T\s{eff} = T + \Delta T$.
We estimate order change under illumination by taking shift along the $Q(T)$ relation of the dark state.
We can examine the effect of \textit{cis} impurities on the free energy.
The nematic mean field potential is $U(\theta) = - JQP_2(\cos(\theta))$, where $\theta$ is the angle
a rod-like mesogen makes with the nematic director, $J$ is the nematic mean field coupling constant, and $P_2$ is the second-order
Legendre polynomial. The coupling constant $J$ depends on the concentration $\rho$ of linear rods as $\rho^2$~\cite{warnerbook:07}.
We assume for simplicity that bent \textit{cis} isomers have no nematic order; they weaken the effective potential experienced by
the linear rods. One can show that the temperature-dependent part of Landau -- de Gennes expansion of the
Maier-Saupe~\cite{chandrasekhar:92} free energy has the form $F\s{L} = \frac{1}{2}A_0 (T - T^*)Q^2 + \dots$, where $T^* = J/(5k\s{B})$ is the transition temperature.
Upon illumination, one can view the change of $\rho$, and thus change of $J$ and $T^*$, as an effective change in $T$ at constant $T^*$.

Our nematic photo-elastomers contain regular mesogenic molecules and azo dyes that contribute to the nematic
order when in the \textit{trans} state. Let $\rho_0$ denote the total concentration of all mesogenic molecules
in the dark state (no \textit{cis} molecules present), and $\delta$ the molar fraction of azo dyes. Upon illumination, the
total concentration of linear molecules after time $t$ is $\rho(t) = \rho\s{h} + \rho\s{t}(t)$, where $\rho\s{h}= \rho_0(1-\delta)$
is the concentration of regular mesogenic molecules (constant in time), and $\rho\s{t}$ is that of azo dyes in
the \textit{trans} state at time $t$ ($\rho\s{t}(t=0) = \rho_0 \delta $). Dye molecules in \textit{trans} and \textit{cis} states
contribute to the total dye molecules concentration: $\rho_0 \delta = \rho\s{t}(t) + \rho\s{c}(t)$, where $\rho\s{c}(t)$ is the concentration of
\textit{cis} molecules at time $t$. Expressed in terms of \textit{trans} $n\s{t} = \rho\s{t}/\rho_0 \delta$ and \textit{cis}
$n\s{c} = \rho\s{c}/\rho_0 \delta$ number fractions, the previous relation becomes $n\s{t}(t) + n\s{c}(t) = 1$.
For the total concentration of linear molecules $\rho(t)$ one can write
$\rho(t) = \rho_0 [1 - \delta + n\s{t}(t) \delta] = \rho_0 [1 - n\s{c}(t) \delta]$.
Now, the coupling constant $J$ becomes $J(t) = J_0 [1 - n\s{c}(t) \delta]^2 \approx J_0 [1 - 2 n\s{c}(t) \delta]$
for $n\s{c}(t) \delta \ll 1$. Thus, \textit{cis} impurities decrease $T^*$ by $\Delta T = 2 T^* n\s{c} \delta$, which is
equivalent to increasing the real temperature to $T\s{eff} = T + \Delta T = T + 2 T^* n\s{c} \delta$.

The actual order parameter (upon illumination) at temperature $T$ is then the dark state order parameter
$Q(T)$ shifted to $Q(T\s{eff}) = Q(T + \Delta T) = Q(T + 2 T^* n\s{c} \delta)$.
If $T$ is not too close to the transition temperature, $T\s{ni}$, the last expression can be
approximated by~\cite{cviklinski:02} $Q(T\s{eff}) \approx Q(T) + b n\s{c}$, where we have introduced $b \equiv 2 (dQ/dT) T^* \delta<0$,
since $(dQ/dT) < 0$. Linearization leads to a simple stress off-dynamics $\overline{\sigma}\s{n}(t) = {\rm e}^{-t/\tau}$ (to be discussed in Section \ref{subsect:B}) which is compatible with experimental findings. The exact expression for $Q(T\s{eff})$ (without linearization)
converts the simple exponential time decay of $n\s{c}$ to a non-exponential time response of $\overline{\sigma}\s{n}$.
On substituting $\lambda\s{ps}$ in Eq.~(\ref{stress}) and writing the
elements of step length tensors $\tens{\ell}\s{0}$ and $\tens{\ell}\s{p}$ in terms of the order parameters $Q(T)$
and $Q(T\s{eff})= Q(T) + b n\s{c}$, one gets a cumbersome expression for $\sigma$. Keeping only linear terms
in $bn\s{c}$, consistently with linear decay processes, one obtains
\be
\sigma = -3 b \mu \frac{n\s{c}}{(1-Q(T))(1+2Q(T))}.
\label{stressnclin}
\ee
The stress $\sigma$ is of course positive since $b$ is negative. The \textit{cis} number fraction is time and depth dependent
$n\s{c} = n\s{c}(x,t)$, which leads to a ($x$,$t$) dependence of the stress. To calculate the average stress $\overline{\sigma}(t)$,
that is a measure of the force, exerted on the clamps one must integrate over depth
$\overline{\sigma}(t) = \int_0^d \sigma(x,t) {\rm{d}} x/d$, where $d$ is the sample thickness. To do this, we have to explore the
behavior of the \textit{cis} number fraction $n\s{c}(x,t)$.

\subsection{Light absorption}\label{subsect:B}

The light intensity $I$ varies with depth $x$ (see Fig.~\ref{fig1}) due to absorption by
\textit{trans} and \textit{cis} species of dye molecules
\be
\frac{\partial I}{\partial x} = - \gamma \Gamma\s{t} I(x,t) n\s{t}(x,t) - \gamma \Gamma\s{c} I(x,t) n\s{c}(x,t),
\label{intensity}
\ee
where $\Gamma\s{t}$ and $\Gamma\s{c}$ are rate coefficients for photon absorption. Here $\gamma = \hbar
\omega \rho_0 \delta$ subsumes the energy $\hbar \omega$ of each absorption of a photon from the beam and the absolute
number density of chromophores $\rho_0 \delta$.
To simplify the above relation we normalize $I(x,t)$ by the incident intensity $I_0$ to obtain an $\mathcal{I} = I/I_0$.
The combinations $\gamma \Gamma\s{t}$ and $\gamma \Gamma\s{c}$ will be written as $1/d\s{t}$ and $1/d\s{c}$ respectively.
Now Eq.~(\ref{intensity}) becomes
\be
\frac{\partial \mathcal{I}}{\partial x} = - \frac{n\s{t}(x,t)}{d\s{t}} \mathcal{I}(x,t) - \frac{n\s{c}(x,t)}{d\s{c}} \mathcal{I}(x,t).
\label{intensityr}
\ee
Assuming that the the \textit{trans}-population of absorbers does not change appreciably, $n\s{t} (x,t) \simeq 1$,
one obtains Beer attenuation $\mathcal{I} = \e^{-x/d\s{t}}$. To close Eq.~(\ref{intensityr}) one needs the rate
equation for the \textit{trans}-population at ($x$,$t$),
\be
\frac{\partial n\s{t}}{\partial t} = - \eta\s{t} \Gamma\s{t} I(x,t) n\s{t}(x,t) + \left ( \eta\s{c} \Gamma\s{c} I(x,t) +
\frac{1}{\tau} \right ) n\s{c}(x,t).
\label{transnumber}
\ee
The changes in $n\s{t}$ are due to photoconversions (with quantum efficiencies
$\eta\s{t}$ per photon absorption of \textit{trans} to \textit{cis} transition, and $\eta\s{c}$ per photon absorption of
\textit{cis} to \textit{trans} transition), and thermal back reaction from the
\textit{cis}-population at a rate $1/\tau$, with $\tau$ being the \textit{cis} lifetime. Eq.~(\ref{transnumber}) can be rewritten
as
\be
\tau \frac{\partial n\s{t}}{\partial t} = (1 + \beta \mathcal{I}(x,t)) - [1+(\alpha + \beta)\mathcal{I}(x,t)]n\s{t}(x,t),
\label{transnumberab}
\ee
where the combinations $\alpha = \eta\s{t} \Gamma\s{t} I_0 \tau$ and
$\beta = \eta\s{c} \Gamma\s{c} I_0 \tau$ measure how intense the incident beam is compared with the material constants
$I\s{t} \equiv 1/(\eta\s{t} \Gamma\s{t} \tau)$ and $I\s{c} \equiv 1/(\eta\s{c} \Gamma\s{c} \tau)$.
The parameter $\alpha$ is the ratio between the forward \textit{trans} to \textit{cis} conversion rate $\eta\s{t} \Gamma\s{t} I_0$,
and the thermal back rate, $1/\tau$; parameter $\beta$ has similar interpretation.
Note that both $\alpha$ and $\beta$ depend on temperature $T$ since the \textit{cis} to \textit{trans} thermal decay
is activated, $\tau = \tau(T)$, and $\tau$ is measurable directly or in stress decay.

After integration of Eq.~(\ref{intensityr}) over depth $x$ we get
\be
\int_0^d n\s{c}(x,t) {\rm{d}} x = \frac{d-d\s{t} \mathcal{A}(d,t)}{1-d\s{t}/d\s{c}},
\label{intcis}
\ee
where the absorption is determined by the relation $\mathcal{A}(x,t) = - \ln \mathcal{I}(x,t)$.
From the above equation one easily obtains the average stress
\be
\overline{\sigma}(t) = - \frac{3b\mu}{(1-Q)(1+2Q)} \frac{1-(d\s{t}/d)\mathcal{A}(d,t)}{1-d\s{t}/d\s{c}}.
\label{stressnonnorm}
\ee
In experiments one usually measures the normalized stress
$\overline{\sigma}\s{n}(t) = \overline{\sigma}(t)/\overline{\sigma}\s{s}$, where
$\overline{\sigma}\s{s}$ is the stationary state stress,
\be
\overline{\sigma}\s{n}(t) = \frac{1-(d\s{t}/d)\mathcal{A}(d,t)}{1-(d\s{t}/d)\mathcal{A}(d)}.
\label{stressnorm}
\ee
Here we adopt the convention that $\mathcal{A}(d)$ denotes the steady state value of the absorption $\mathcal{A}(d,t)$.
The same convention will be used for $\mathcal{I}$, $n\s{t}$ and $n\s{c}$.

For off-dynamics, after setting $I=0$ in Eq.~(\ref{transnumber}), one gets a simple equation for the \textit{cis}
number fraction $n\s{c}(x,t)= n\s{c}(x,0) {\rm e}^{-t/\tau}$, where $n\s{c}(x,0)$ represents the spatial profile of $n\s{c}$
at the instant the light is switched off $t=0$. On inserting $n\s{c}(x,t)$ into Eq.~(\ref{stressnclin}),
averaging over depth, and normalizing the average stress with its value at $t=0$, one gets
a simple expression $\overline{\sigma}\s{n}(t) = {\rm e}^{-t/\tau}$.

\begin{figure}
  \centering
  \includegraphics[width=8.7cm]{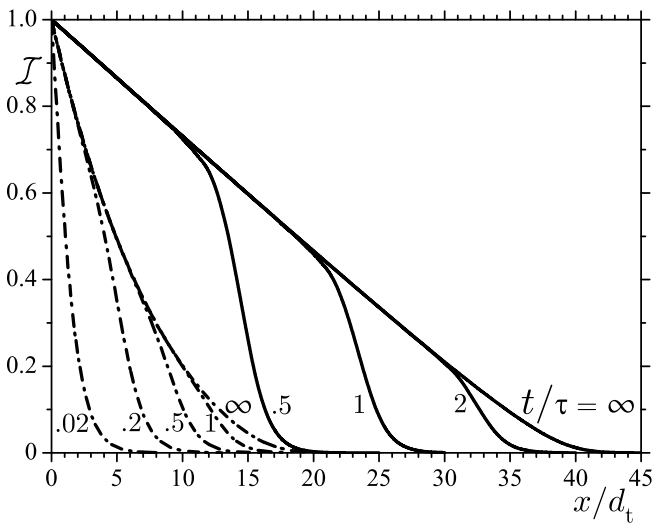}
  \caption{Light intensity versus the reduced depth $x/d\s{t}$ for various reduced times $t/\tau$.
  The dash-dotted lines represent the case with the presence of back photoconversion ($\alpha = 36$, $\beta =1$, and $\eta=3$ in this example),
  while the solid lines correspond to the absence of back photoconversion ($\alpha = 36$, $\beta =0$); note that even a small
  value of $\beta$ (compared to $\alpha$) significantly affects the variation of $\mathcal{I}$ with depth.
  In the limit of large $t/\tau$ one achieves the stationary regime described by Eq.~(\ref{eqsolgen}).
  }
\label{fig2}
\end{figure}

Given that the stress $\overline{\sigma}\s{n}(t)$ for on-dynamics depends on both $\mathcal{A}(d,t)$ and $\mathcal{A}(d)$, we
shall firstly discuss the behavior of these quantities.
In the steady state, $\partial n\s{t}/ \partial t = 0$, we have
\be
n\s{t}(x) = \frac{1+ \beta \mathcal{I}(x)}{1 + (\alpha + \beta) \mathcal{I}(x)}, \quad n\s{c}(x) =
\frac{\alpha \mathcal{I}(x)}{1 + (\alpha + \beta) \mathcal{I}(x)}.
\label{transeq}
\ee
Now the equilibrium solution of Eq.~(\ref{intensityr}) can be expressed in the form
\be
\ln(\mathcal{I}) + \left ( \frac{\alpha - \eta \beta}{\beta(1+\eta)} \right )
\ln \left ( \frac{1+\beta(1+\eta)\mathcal{I}}{1+\beta(1+\eta)} \right ) = - \frac{x}{d\s{t}},
\label{eqsolgen}
\ee
where $\eta = \eta\s{t}/\eta\s{c}$ is the ratio of the quantum efficiencies. The above expression for $\mathcal{I}(x)$
provides a generalization of the usual Beer's law.
Deviations from Beer's law come about because at high intensities the \textit{cis} population increases and is
generally less absorbing than the \textit{trans} species. In the limit $\Gamma\s{c} \rightarrow 0$, the parameter
$\beta \rightarrow 0$ so that Eq.~(\ref{eqsolgen}) reduces to
\be
\ln (\mathcal{I}) + \alpha (\mathcal{I} - 1) = - \frac{x}{d\s{t}}.
\label{eqsol}
\ee
For $\alpha =0$, one finds the standard Beer's law $\mathcal{I}={\rm e}^{-x/d\s{t}}$, while for large $\alpha$
the law acquires the linear form
\be
\mathcal{I}(x) \simeq 1 - \frac{x}{\alpha d\s{t}},
\label{connonlin}
\ee
at least over depths up to $x \sim \alpha d\s{t}$ whereupon $\mathcal{I}$ is
small and the $\ln (\mathcal{I})$ again prevails to give a finally exponential penetration.
The variation of light intensity with reduced depth for large $\alpha$ is represented in Fig.~\ref{fig2}, by the solid line $t/\tau = \infty$ in the case of absence of back photoconversion ($\beta =0$), and by the dash-dotted line $t/\tau = \infty$ when $\beta \ne 0$. Note that even a moderate value of $\beta$ (see Fig.~\ref{fig2}) has a major impact on the variation of $\mathcal{I}$ with depth.
Non-Beer absorption was first explored for dyes in nematic liquid crystals by Statman and Janossy~\cite{statman:03} and by Corbett and Warner~\cite{corbettdyn:03,corbettdyn:02,corbettdyn:01},
and experimentally by Serra and Terentjev~\cite{serra:08}.

Beer absorption has no dynamics since it holds only if the number fraction $n\s{t}$ is unchanging, $n\s{t} =1$.
To explore the dynamics of non-Beer absorption, we have to solve the coupled
Eqs.~(\ref{intensityr}) and (\ref{transnumberab}). Indeed, using these two equations one gets~\cite{corbettdyn:01}
\be
\tau \frac{\partial \mathcal{A}}{\partial t} = - \mathcal{A} + \frac{x}{d\s{t}} + (\alpha + \beta)(\mathcal{I} - 1) +
\left ( \frac{\alpha}{d\s{c}} + \frac{\beta}{d\s{t}} \right ) \int_0^x \mathcal{I} {\rm d} x.
\label{abseqn}
\ee
Clearly, the absorption $\mathcal{A}$ can be expressed in terms of reduced variables $\tilde{x} = x/d\s{t}$ and $\tilde{t} = t/\tau$.
Taking derivative of the above equation with respect to $\tilde{x}$, and using $\mathcal{I} = {\rm e}^{-\mathcal{A}}$, we get
a partial differential equation in $\mathcal{A}(\tilde{x}, \tilde{t})$,
\be
\frac{\partial^2 \mathcal{A}}{\partial \tilde{t} \partial \tilde{x}} = 1 - \frac{\partial \mathcal{A}}{\partial \tilde{x}} \left [ 1 + (\alpha + \beta) {\rm e}^{-\mathcal{A}} \right ] + \beta(1+\eta){\rm e}^{-\mathcal{A}}.
\label{absorption}
\ee
We analyze this equation numerically using the boundary condition $\mathcal{A}(0,\tilde{t}) = 0$ and the initial condition $\mathcal{A}(\tilde{x},0) = \tilde{x}$. The results are presented in Fig.~\ref{fig2}. As expected the light intensity decreases with reduced depth more rapidly in the presence of back photoconversion ($\beta \ne 0$, dash-dotted lines) than in the absence of back photoconversion ($\beta = 0$, solid lines).
Note that in both cases the stationary regime is reached for moderate values of the reduced time $t/\tau$.
\begin{figure}
  \centering
  \includegraphics[width=8.7cm]{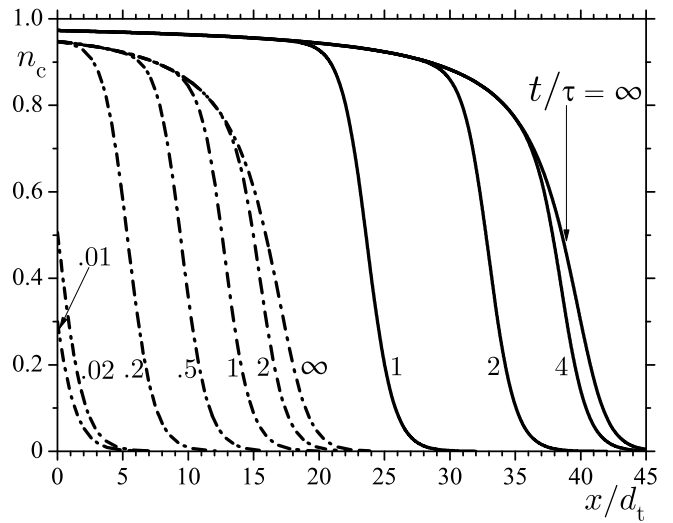}
  \caption{The \textit{cis} number fraction $n\s{c}$ against the reduced depth $x/d\s{t}$ for various reduced times $t/\tau$.
  The dash-dotted lines represent the case $\alpha = 36$, $\beta =1$, and $\eta=3$ while the solid lines represent the case
  $\alpha = 36$, $\beta =0$. In the large time limit one approaches the solution presented by the second equation of~(\ref{transeq}).}
\label{fig3}
\end{figure}

Before proceeding with the analysis of the behavior of the stress, it is tempting to consider briefly the time evolution of the
\textit{cis} number fraction $n\s{c} = n\s{c}(\tilde{x},\tilde{t})$. Thus, by taking $\mathcal{I}$ from Eq.~(\ref{transnumberab})
and substituting it into Eq.~(\ref{intensityr}), one gets
\bea
&(&\dot{n}\s{c} + n\s{c}) \left [ 1 - n\s{c} \left ( 1 + \frac{\beta}{\alpha} \right ) \right ]
\left [ 1 - n\s{c} \left ( 1 - \frac{\eta \beta}{\alpha} \right ) \right ] \nonumber \\
= &[& \dot{n}\s{c}' (n\s{c} - 1) - n\s{c}' (1 + \dot{n}\s{c}) ] +
\frac{\beta}{\alpha} \left ( \dot{n}\s{c}' n\s{c} - \dot{n}\s{c} n\s{c}' \right ),
\label{ncpde}
\eea
where for simplicity we used the notation $\dot{n}\s{c} = \partial n\s{c}/\partial \tilde{t}$ and
$n\s{c}' = \partial n\s{c}/\partial \tilde{x}$.
In the numerical integration of this equation we use the boundary condition
\be
n\s{c}(0,\tilde{t}) = \frac{\alpha}{1 + \alpha + \beta}
\{1 - \exp [-\tilde{t} (1+ \alpha + \beta)]\},
\label{ncpdebc}
\ee
and the initial condition $n\s{c}(\tilde{x},0) = 0$. The quoted boundary condition is obtained by integration
of Eq.~(\ref{transnumberab}) for $x=0$. Figure~\ref{fig3} shows $n\s{c}$ as a function of the reduced depth $x/d\s{t}$ for
various reduced times $t/\tau$. It seems that the stationary solution, given by the second equation of~(\ref{transeq}), is reached already
for moderate values of the reduced time $t/\tau$.

\begin{figure}
  \centering
  \includegraphics[width=8.7cm]{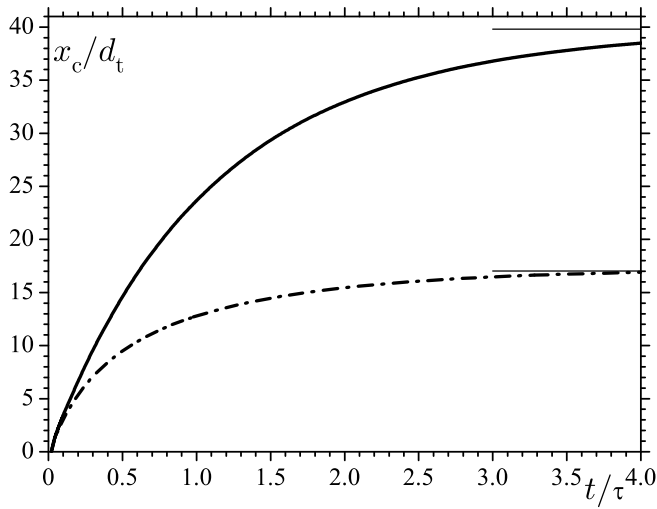}
  \caption{Approximate thickness $x\s{c}/d\s{t}$ of \textit{cis} layer against the reduced time $t/\tau$ for
  $\alpha=36$, $\beta=0$ (solid line) and $\alpha=36$, $\beta=1$, and $\eta =3$ (dash-dotted).
  Both curves reach their stationary values after a moderate time (measured in units of $\tau$)
  which coincide with corresponding values (thin solid lines) obtained from Eq.~(\ref{inflex}) as explained in the main text.
  }
\label{fig4}
\end{figure}

We estimate the thickness $x\s{c}/d\s{t}$ of \textit{cis} layer at $t/\tau$ by calculating positions of points
of the curves from Fig.~\ref{fig3} where second derivatives of $n\s{c}$ over $x/d\s{t}$ change the sign.
In the stationary case, this condition can be expressed analytically, by taking second derivative of the second equation of (\ref{transeq}) with respect to $x/d\s{t}$ and using (\ref{intensityr}),
\be
(\alpha + \beta) \beta (1 + \eta) \mathcal{I}\s{c}^2 + 2 (\alpha - \eta \beta) \mathcal{I}\s{c} - 1 = 0,
\label{inflex}
\ee
where $\mathcal{I}\s{c} = \mathcal{I}(x\s{c}/d\s{t})$. Inserting the positive solution of this equation into Eq.~(\ref{eqsolgen}) we obtain an equation for the stationary value $x\s{c}/d\s{t}$.
In the non-stationary case we proceed numerically. Results for $\alpha=36$, $\beta=0$, and $\alpha=36$, $\beta=1$ are presented in Fig.~\ref{fig4}. In both cases $x\s{c}/d\s{t}$ increases approximately
logarithmically with time in the interval from about $0.5 \tau$ to $1.5 \tau$; this can be easily seen
if one presents curves from Fig.~\ref{fig4} on a logarithmic scale.
The stationary solution is reached in about $t = 4 \tau$ for $\beta =1$ and $t= 5 \tau$ for $\beta=0$. It is also interesting to compare the time needed to reach the stationary state with finite $\beta$ with the time needed to convert equally thick layer without \textit{cis} absorption. In Fig.~\ref{fig4} this ratio is about $10$, so even though the \textit{cis}
absorption is small, it strongly affects the photodynamics of the system.
The back photoconversion has also been found to be important in the analysis of experimental
findings~\cite{copic:11}.

\section{Results}\label{sect:results}

As we have seen, the absorption $\mathcal{A}$ at the back of the sample, $x = d$, in Eq.~(\ref{stressnorm}) for stress
is actually a function of reduced variables, $\mathcal{A}(\tilde{d},\tilde{t})$.
Then one can generate $\overline{\sigma}\s{n}(t)$ for different values of parameters $\alpha$, $\beta$, $\tilde{d}$ and $\tau$.
Experimental data for $\overline{\sigma}\s{n}(t)$ was fitted~\cite{ferrer:11}
to the simple empirical form $\overline{\sigma}\s{n}(t) = 1 - \exp [-(t/\tau\s{on})^{\beta\s{on}}]$, with $\beta\s{on}<1$.
This stretched exponential form must fail at short times, and fitting at long times is difficult. We have fitted our theoretical $\overline{\sigma}\s{n}(t)$ to the above form. In the absence of \textit{cis} to \textit{trans} photoconversion ($\beta = 0$) we can
get fits for $\beta\s{on}>1$ only. If we allow, however, back photoconversion we can obtain agreement with the stretched exponential form, $\beta\s{on}<1$. It is, therefore, very important to take into account back photoconversion to successfully fit the experimental data.

We consider the experimental data for compound SCEAzo2-c-10 in Fig.~5(a) of Ref.~\cite{ferrer:11}. There are several quantities entering our relation (\ref{stressnorm}) for stress: the ratio of the quantum efficiencies $\eta$, the characteristic thermal relaxation time $\tau$, thickness of the sample $\tilde{d}= d/d\s{t}$ in units of the characteristic length, and the dimensionless parameters $\alpha$ and $\beta$.
For the ratio of the quantum efficiencies we adopt the estimate $\eta \approx 3$ of Ref.~\cite{ikedabook:09}; some larger values
of $\eta$ (up to $\eta \approx 4$) have also been reported. Our analysis shows, however, that the quality of the fit is not very much affected by the particular value of $\eta$ in the range $\eta \approx 3 - 4$. For the thermal relaxation times $\tau(T)$ we take the experimental values
found in the stress off-dynamics~\cite{ferrer:11}: $\tau= 51.5,~ 33,~ 23,~ 15,~ 10.4,~ 7.2 ~{\rm min}$, for temperatures
$T= 45,~ 50,~ 55,~ 60,~ 65,~ 70 ~{}^\circ {\rm C}$ respectively. With this choice the number of required fit parameters
is reduced to three, $\tilde{d}$, $\alpha$, and $\beta$.
\begin{figure}
  \centering
  \includegraphics[width=8.7cm]{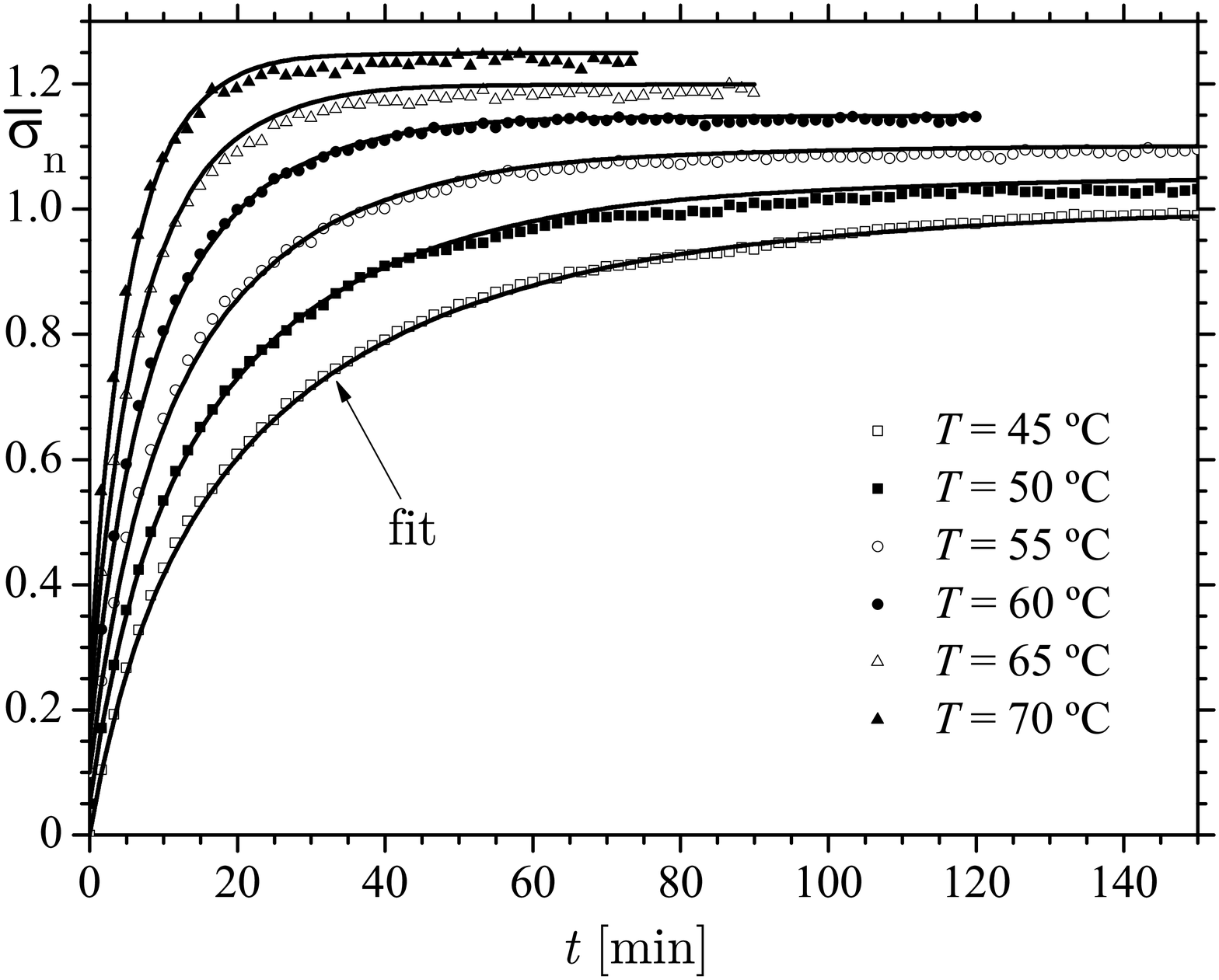}
  \caption{Normalized stress as a function of time. Theoretical predictions are presented by solid lines, while experimental data, taken from~\cite{ferrer:11}, are presented by different symbols corresponding to different temperatures. For better visibility the five curves corresponding
  to temperatures $T>45~{}^\circ {\rm C}$ are shifted along the $\overline{\sigma}\s{n}$-axis by the values
  $0.05, ~0.1, ~0.15, ~0.2, ~0.25$ respectively. Only the $T_0 = 45~{}^\circ {\rm C}$ curve is fitted, while all other curves
  are theoretical.
  }
\label{fig5}
\end{figure}

We first fit the experimental data for $\overline{\sigma}\s{n}(t)$ at $T_0 = 45 ~{}^\circ {\rm C}$ to our expression~(\ref{stressnorm}),
using $\tilde{d}$, $\alpha$ and $\beta$ as fit parameters; see Fig.~\ref{fig5}.
The optimal values of fit parameters were found to be $\alpha= 54.9$, $\beta = 1.7$, and $\tilde{d}=20$.
Given that the sample thickness~\cite{ferrer:11} was $d=300 ~\mu{\rm m}$, the corresponding Beer length
is $d\s{t} = 15~\mu{\rm m}$, which is not far from an independent estimate $d\s{t} \approx 10~\mu{\rm m}$ obtained from the azobenzene absorption spectra. This rough estimate of $d\s{t}$ can be obtained by assuming that the molar extinction for SCEAzo2-c-10 at its absorption maximum coincides with the corresponding known value for the azobenzene in benzene~\cite{note:1}.
Let us note, however, that quite good fits can be also obtained for some other values of fit parameters. This ambiguity can be settled obviously by a further reduction of the number of the fit parameters -- for example by measuring $\tilde{d}$.
Anyway, in our case situation is not so serious, because the parameter $\tilde{d}$ is temperature independent, while parameters $\alpha$ and $\beta$ depend on temperature $T$ only through thermal relaxation time,
$\alpha(T) = \eta\s{t} \Gamma\s{t} I_0 \tau(T)$ and $\beta(T) = \eta\s{c} \Gamma\s{c} I_0 \tau(T)$.
For example, at other temperatures $T$, one can write $\alpha(T) = \alpha(T_0)\tau(T)/\tau(T_0)$ and
$\beta(T) = \beta(T_0)\tau(T)/\tau(T_0)$. Thus taking the experimental values of thermal relaxation times $\tau(T)$ for
temperatures quoted in Fig.~\ref{fig5} we determine the corresponding values of $\alpha(T)$ and $\beta(T)$. Then
using these values and expression~(\ref{stressnorm}) for the normalized stress we simply plot corresponding curves
for temperatures $T>T_0$ without any fitting. Note excellent agreement of these theoretical predictions with experimental
data. Extracting $\alpha$ and $\beta$ by fitting at one $T_0$ gives universal, fit-free agreement at other temperatures.

\section{Conclusions}\label{sect:conclusions}

We have demonstrated that both significant force magnitude and complex force dynamics result from nonlinear optical absorption.
A bleaching wave of increased \textit{cis} concentration, and hence contribution to retractile force, penetrates a
sample with a highly characteristic dynamics in which the (small) absorption of the \textit{cis}
moiety is essential. Fitting $\overline{\sigma}\s{n}(t)$ at one temperature yields the
relevant material parameters of the dye responsible for the opto-mechanical response. The $\overline{\sigma}\s{n}(t)$
response at other temperatures is then obtained with no further fit by simply scaling these parameters by the
separately measured decay rates at the other temperatures. This astonishing predictive power points to the
validity of the nonlinear temporal-spatial optical absorption model used.

Future experiments should separately measure Beer penetration depths in the weak absorption limit, and the
nonlinear material parameters. In that event there should be no fit parameters at all, as in this current
work away from the reference temperature. With the material parameters thus measured, one could employ the theory
to examine more complex systems such as non-clamped elastomers and systems where the response is more complex due to
director patterning.

\section*{Acknowledgments}

M. K. thanks the Winton Programme for the Physics of Sustainability and the Cambridge Overseas Trust for financial support, and M. W. thanks
the Engineering and Physical Sciences Research Council (UK) for a Senior Fellowship. M. \v{C}. thanks the Winton Programme and Isaac Newton Institute for support.
We thank Eugene Terentjev for useful discussions.

\end{document}